\documentclass[twocolumn,prl,aps]{revtex4}
\usepackage{graphics}
\begin{document}
\title{Electron-phonon interaction in the $t$-$J$ model}
\author{O. R\"osch  and O. Gunnarsson}
\affiliation{ Max-Planck-Institut f\"ur Festk\"orperforschung, 
D-70506 Stuttgart, Germany}

\begin{abstract}
We derive a $t$-$J$ model with electron-phonon coupling from the three-band 
model, considering modulation of both hopping and Coulomb 
integrals by phonons. While the modulation of the hopping integrals
dominates, the modulation of the Coulomb integrals cannot be neglected.
The model explains the experimentally observed anomalous softening of the 
half-breathing mode upon doping and a weaker softening of the 
breathing mode. It is shown that other phonons are not strongly influenced,
and, in particular, the coupling to a buckling mode is not strong in 
this model.
\end{abstract}

\maketitle

There has been a strong interest in electron-phonon coupling
for high-$T_c$ cuprates after the discovery of strong coupling
to a mode at 70 meV in many cuprates \cite{Lanzara}. The coupling 
was ascribed to a half-breathing phonon along the (1,0,0) direction, 
i.e., an in-plane bond-stretching mode, with an electron-phonon 
coupling $\lambda \sim 1$. This phonon shows an  anomalous softening 
when the cuprates are doped, in particular, towards the zone 
boundary \cite{Pintschovius1,Pintschovius2,Pintschovius3,anomol}.  
While the softening of other phonons upon doping can be understood 
as a screening of the ions, the softening of the half-breathing mode 
cannot be described in a shell model with conventional 
parameters \cite{Pintschovius1}. This suggests a substantial 
electron-phonon coupling.  The phonon has an appreciable 
broadening \cite{Pintschovius2}, also supporting a substantial coupling. 
Local-density approximation
(LDA) calculations, however, show a weak coupling to this 
phonon, with $\lambda$ at the zone boundary being $\sim$ 
0.01 \cite{Savrasov,Bohnen}. Anomalous behavior of bond-stretching modes has 
also been observed in other compounds \cite{Tranquada,Braden,Reichardt}. It is 
then interesting to study the electron-phonon interaction, taking 
into account the strong electron-electron interaction \cite{Dagotto,Scalapino}. 
This has been done in the $t$-$J$ model \cite{Zhang} by von Szczepanski and 
Becker \cite{Becker}, by Khaliullin and Horsch \cite{Horsch}
and by Ishihara, Nagaosa and coworkers \cite{Nagaosa}. 

The $t$-$J$ model is derived from a three-band model of a CuO$_2$ 
layer \cite{Emery}, including a Cu $d_{x^2-y^2}$ and two 
O $p$ orbitals. To obtain the electron-phonon coupling, the atoms 
are displaced. This leads to a change of the hopping integral 
$t_{pd}$ and the charge transfer energy $\varepsilon_p$ between 
Cu and O atoms, where the change of $\varepsilon_p$ is assumed to
be due to a change of the Coulomb integral $U_{pd}$ between nearest 
neighbor Cu and O atoms. Transforming to a   $t$-$J$ 
model, this leads to changes of both the on-site energy $E_s$ of the 
Zhang-Rice singlet and the hopping between the sites (off-site coupling). 
Two groups \cite{Becker,Horsch} assumed the variation of $t_{pd}$
to dominate, while a third group \cite{Nagaosa} focused on    
$\varepsilon_p$. The first two groups assumed the off-site 
coupling in the $t$-$J$ model to be negligible, while the third group 
emphasized this coupling. 

The singlet energy in the $t$-$J$ model is large, $|E_s|\sim 5$ eV.
For a rigid lattice and fixed doping, this energy enters only as an
uninteresting constant. Because of the strong distance dependence
of $t_{pd}$, however, we may expect a strong electron-phonon
coupling from this term.

We study modulations of $t_{pd}$ and $\varepsilon_p$ 
by phonons. The $t_{pd}$ modulation dominates, but destructive 
interference between the two effects is not negligible.   The phonons 
couple mainly to on-site terms. 
Phonon spectral functions are obtained from exact diagonalization.
The model explains the anomalous 
softening of the half-breathing mode and correctly gives  a weaker 
softening of  the breathing mode. Other phonons are only softened 
weakly. Experiments suggest a strong coupling of preferentially 
the anti-nodal electronic state to a phonon at 40 meV, perhaps a 
$B_{1g}$ buckling mode \cite{Shen}. We find a weak coupling to the 
$B_{1g}$ mode in the model studied here.

The derivation of a model with phonons proceeds as for the normal $t$-$J$ 
model \cite{Zhang}, but with displaced 
atoms. With the atoms in their equilibrium positions, a Cu atom
couples only to a given linear combination of $p$-orbitals on the four O 
neighbors \cite{Zhang}. With displaced atoms there 
is also coupling to other combinations. This coupling, however, 
enters to second order in the displacement and is neglected here.
Following Zhang and Rice \cite{Zhang}, we work to lowest order in
$t_{pd}$, although this is not very accurate for 
realistic parameters. We neglect certain terms where a hole     
hops $i\to k \to j$ ($k\ne i$,  $k\ne j$). We also neglect the 
electron-phonon interaction via superexchange interaction,
since this coupling occurs only via motion of Cu atoms and with
a small prefactor \cite{Becker}. We then obtain a $t$-$J$ model 
with the linear electron-phonon coupling            

\begin{equation}\label{eq:1}
H_{\mathrm{el-ph}}=\sum_{ij\sigma}\tilde c_{i\sigma}^{\dagger}\tilde c_{j\sigma}
\sum_{{\bf q},\nu}g_{ij}({\bf q},\nu)(b_{{\bf q},\nu}+
b_{-{\bf q},\nu}^{\dagger}),
\end{equation}
where
\begin{eqnarray}\label{eq:2}
&&\!\!\!\!\!\!\!\!\!\!\!\!\!\!g_{ij}=i\sqrt{{2\hbar\over N\omega_{\nu}({\bf q})}}e^{i{\bf q}\cdot ({\bf R}_i+
{\bf R}_j)/2}\lbrack A({\bf q},\nu)\delta_{ij}+B^x({\bf q},\nu)
\delta_{i,j\pm \hat x}
\nonumber \\
&&\quad\quad\quad\quad\quad\quad\quad\quad\quad\quad\quad
+B^y({\bf q},\nu)\delta_{i,j\pm \hat y}\rbrack.
\end{eqnarray}
Here $\tilde c_{i\sigma}^{\dagger}$ creates a $d$-hole on site $i$ if this
site previously had no hole, and $b_{{\bf q},\nu}^{\dagger}$ creates a 
phonon with wave vector ${\bf q}$, index $\nu$ and frequency $\omega_{\nu}
({\bf q})$. The system has $N$ sites with coordinates ${\bf R}_i$.
$\delta_{i,j-\hat x}=1$ if site $j$ is one site to the right
of $i$ in the $x$-direction.
The modulation of $t_{pd}$ gives                    
\begin{eqnarray}\label{eq:3}
A\!=\!2 t_{pd} {dt_{pd}\over dr}\left({2 \lambda^2 \!-\!1\over
\varepsilon_p}\! +\! {2 \lambda^2 \over U\!-\!\varepsilon_p}\right)
\!\!\left[{\epsilon_{\mathrm{O}x}^x\over \sqrt{M_\mathrm{O}}} s_x\!+\!
{\epsilon_{\mathrm{O}y}^y\over \sqrt{M_{\mathrm{O}}}} s_y\right]
\end{eqnarray}
and
\begin{eqnarray}\label{eq:4}
&&\!\!\!\!\!\!\!\!\!\!B^x={\lambda \over 4} t_{pd}
{dt_{pd}\over dr}\left({1\over 2}{1\over \varepsilon_p}+
{1\over U-\varepsilon_p}\right)
\nonumber \\
&&
\!\!\!\!\!\!\!\!\times\!\left[ \Gamma_1{\epsilon_{\mathrm{Cu}}^x
\over \sqrt{M_{\mathrm{Cu}}}}  s_x   +2\Gamma_2{\epsilon_{\mathrm{O}x}^x\over
\sqrt{M_\mathrm{O}}}s_xc_x
+2\Gamma_3{\epsilon_{\mathrm{O}y}^y\over
\sqrt{M_\mathrm{O}}}c_x s_y                                       \right]\!\!,
\end{eqnarray}
with an analogous form for $B^y$.
Here $s_l={\rm sin} (q_la/2)$, $c_l={\rm cos} 
(q_la/2)$, $l=x$, $y$; $\lambda=\sum \beta_{\bf k}^{-1}/N
=\gamma_0-\gamma_1=0.96$,
$\Gamma_1=4(\gamma_2-\gamma_0)$, $\Gamma_2=8\gamma_1-\gamma_0-
5\gamma_2-2\gamma_{11}$ and $\Gamma_3=8\gamma_1-\gamma_0-\gamma_2-
6\gamma_{11}$, with  $\gamma_0=\gamma(0)=1.285$, $\gamma_1=
\gamma(a{\bf \hat x})=0.327$, 
$\gamma_{11}= \gamma(a{\bf \hat x}+a{\bf \hat y})=0.209$ and $\gamma_2=
\gamma(2a{\bf \hat x})=0.164$, where 
$\gamma({\bf R})=\sum_{\bf k}\beta_{\bf k}e^{i {\bf k}\cdot 
{\bf R}}/N$ with $\beta_{\bf k}=1/\sqrt{s_x^2+s_y^2}$
and $a$ the lattice parameter.
$M_{\mathrm{Cu}}$ and $M_\mathrm{O}$ are the masses of the Cu and O atoms, respectively,
$\epsilon^x_{\mathrm{O}x}$ is the $x$-component of the polarization vector 
for one of the O atoms (O$_x$) with an analogous meaning of 
$\epsilon^x_{\mathrm{Cu}}$. $U$ is the hole-hole repulsion on Cu.
von Szczepanski and Becker \cite{Becker} calculated $g_{ij}({\bf q},\nu)$
for the ${\bf q}=(1,1)\pi/a$ breathing mode and found $B^x({\bf q},\nu)=0$ 
(since $\epsilon_{\mathrm{Cu}}^x=0$ for this value of ${\bf q}$), as above, 
and a slightly different result for $A({\bf q},\nu)$ due to slightly 
different approximations. Considering a variation of the 
charge transfer energy, assuming that it is driven by a nearest 
neighbor Cu-O Coulomb interaction $U_{pd}$, we find
\begin{eqnarray}\label{eq:5}
&&\!\!\!\!\!\!\!A=-8 t_{pd}^2{dU_{pd}\over dr}
\left( {2\lambda^2-1\over \varepsilon_p^2}-
{2\lambda^2\over (U-\varepsilon_p)^2}\right)
 \\ &&\!\!\!\!\!\times\!\!
\left[ {\epsilon_{\mathrm{O}x}^x\over \sqrt{M_\mathrm{O}}}s_x
+{\epsilon_{\mathrm{O}y}^y\over \sqrt{M_\mathrm{O}}}s_y+{1\over 2}D\right]
-{dU_{pd}\over dr}D,
\nonumber
\end{eqnarray}
here $D=M_{\mathrm{Cu}}^{-1/2}(\epsilon_{\mathrm{Cu}}^xs_xc_x+\epsilon_{\mathrm{Cu}}^ys_yc_y)$,
and
\begin{eqnarray}\label{eq:6}
&&\!\!\!\!\!\!\!B^x=-4\lambda\alpha t_{pd}^2{dU_{pd}\over dr}
\left({1\over \varepsilon_p^2}-{2\over (U-\varepsilon_p)^2}\right) \nonumber\\
&& \times \left[2{\epsilon_{\mathrm{O}x}^x\over \sqrt{M_\mathrm{O}}}s_xc_x +
2{\epsilon_{\mathrm{O}y}^y\over \sqrt{M_\mathrm{O}}}c_x s_y \right. \\
&&\left.\quad\ +{1\over 4}{\epsilon_{\mathrm{Cu}}^x\over \sqrt{M_{\mathrm{Cu}}}}{\rm sin} 
{3q_x a\over 2}
+{1\over 2}{\epsilon_{\mathrm{Cu}}^y\over \sqrt{M_{\mathrm{Cu}}}}c_xs_y
+{1\over 4}{\epsilon_{\mathrm{Cu}}^x\over \sqrt{M_{\mathrm{Cu}}}}s_x\right]. \nonumber
\end{eqnarray}
Here $\alpha=
\sum \beta_{\bf k}^{-1}e^{ia{\bf k}\cdot {\bf \hat x}}/N=-0.14$.

To estimate the relative magnitude of these terms, we use \cite{parameters,t}
$t_{pd}=1.2$ eV, $\varepsilon_p=3$ eV, $U=10$ eV, $U_{pd}=1$ eV, and $a=3.8$ 
\AA.  We assume the distance dependence $t_{pd}\sim r^{-n}$ with $n=3.5$,
 based on LDA calculations \cite{Ove}, and $U_{pd}\sim r^{-1}$. This gives 
the hopping integral -0.47 eV in the $t$-$J$ model, close to values commonly 
used \cite{Prelovsek}. For  ${\bf q}=(q_x,0)$, $\omega=0.07$ eV,
$\epsilon_{\mathrm{O}x}^x=1$ and all other polarization vectors zero, we obtain
(in eV) 
\begin{equation}\label{eq:7}
\sqrt{{2\hbar /     \omega}}A({\bf q})=-0.25s_x, \hskip0.1cm 
\sqrt{{2\hbar /     \omega}}B^x({\bf q})=-0.0032s_xc_x 
\end{equation}
due to modulation of $t_{pd}$ and
\begin{equation}\label{eq:8}
\sqrt{{2\hbar /     \omega}}A({\bf q})=0.029s_x, \hskip0.1cm 
\sqrt{{2\hbar /    \omega}}B^x({\bf q})=-0.0049s_xc_x 
\end{equation}
due to modulation of $U_{pd}$.
This shows that the modulation of $t_{pd}$ dominates over that of 
$U_{pd}$.  This is mainly  due to a stronger power dependence 
($n=3.5$) for $t_{pd}$ than $U_{pd}$ but also due to a partial 
cancellation between the terms proportional to $1/\varepsilon_p^2$ 
and $1/(U-\varepsilon_p)^2$ in the contribution from $dU_{pd}/dr$.  
The contribution from $dU_{pd}/dr$ alone is very small. There is a 
destructive interference between contributions from $dt_{pd}/dr$
and $dU_{pd}/dr$, however, which reduces the phonon softening by about 
30$\%$. The prefactors of the off-diagonal terms are  small [Eqs. (\ref{eq:7}) and (\ref{eq:8})],      
and the  diagonal term dominates. 
We have neglected quadratic terms in the phonon displacement in 
(\ref{eq:1}), although some terms may give non-negligible contributions 
to the doping dependence of phonon energies.

The model (\ref{eq:1}) describes the softening of phonons due to 
holes in the doped system, but it does not include other interactions 
present in both the doped and undoped systems. These interactions 
are described by a two-spring model, fitted to  the phonon frequencies 
in the (1,0) and (1,1) directions of the {\it undoped} system. This 
spring model provides the eigenvectors $\epsilon$ in Eqs. 
(\ref{eq:3})-(\ref{eq:6}).

\begin{figure}[bt]
\centerline{
{\rotatebox{0}{\resizebox{7.7cm}{!}{\includegraphics {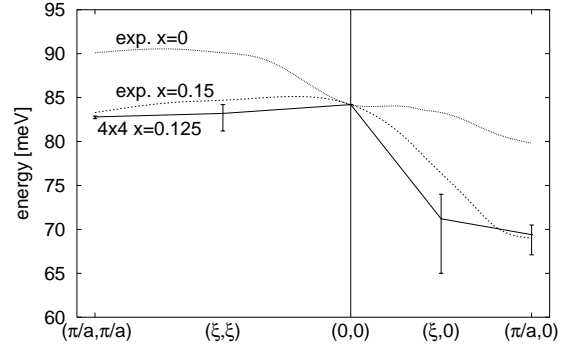}}}}}
\caption[]{\label{fig:1} Phonon dispersion in the (1,0) and (1,1) 
directions. Experimental results (dotted line) for $x=0$ and $x=0.15$
are shown. Theoretical results (full curve) for $x=0.125$ show the    
calculated softening from the experimental $x=0$ 
results. The average over boundary conditions is shown and the bars 
show the spread due to different boundary conditions. There is a strong 
softening in the (1,0) direction, while the softening in the (1,1) 
direction is weaker. 
}
\end{figure}

To study the $t$-$J$ model with phonons, we use exact 
diagonalization \cite{DagottoRMP}, 
including all possible electronic states for a finite cluster of 
size $M\times N$. To obtain a finite Hilbert space, we allow
states containing a maximum of $K(=5)$ phonons \cite{Hilbert}, which is
sufficient for convergence. We calculate the phonon spectral function 
($\omega>0$)
\begin{equation}\label{eq:9}
B_{{\bf q}\nu}(\omega)=-{1\over \pi}{\rm Im}\langle 0 | \phi_{{\bf q}\nu}
{1\over \omega-(H-E_0)+i\eta} \phi_{{\bf q}\nu}^{\dagger}|0\rangle,
\end{equation}
where $|0\rangle$ is the ground-state with the energy $E_0$,
$\phi_{{\bf q}\nu}=b_{{\bf q}\nu}+b^{\dagger}_{-{\bf q}\nu}$
and $\eta$ is infinitesimal. 
Since the clusters that can be treated are too small to give a
quasicontinuous spectrum, we use the center of gravity of the 
phonon spectrum to define the renormalized phonon frequency.
This provides the phonon softening in relation to 
the experimental phonon frequencies of the undoped system.  We 
have used the parameters above and $J/t=0.3$. 

Figure \ref{fig:1} shows results in  a $4\times 4$ $t$-$J$ model for the 
(1,0) and (1,1) directions.  The dotted curves show experimental 
results \cite{Pintschovius1}, and the full curve shows the softening 
due to the electron-phonon interaction in the doped system. Because of
the small cluster sizes, the results depend on the boundary 
conditions. We  have used periodic, antiperiodic, and mixed 
boundary conditions, applying periodic boundary conditions in one 
direction and antiperiodic in the other. Figure \ref{fig:1} shows the 
average and the bars the spread of the results. 
Although the results are fairly sensitive to the boundary conditions, 
the trends are clear. Doping leads to a pronounced softening of the 
half-breathing mode along the (1,0) direction for $|{\bf q}| \gtrsim 
\pi/(2a)$, while the softening is weaker in the (1,1) direction, in 
agreement with experiment \cite{Pintschovius1}. Similar conclusions 
have been obtained using analytical treatments \cite{Horsch}. 

\begin{figure}[bt]
\centerline{
{\rotatebox{0}{\resizebox{7.7cm}{!}{\includegraphics {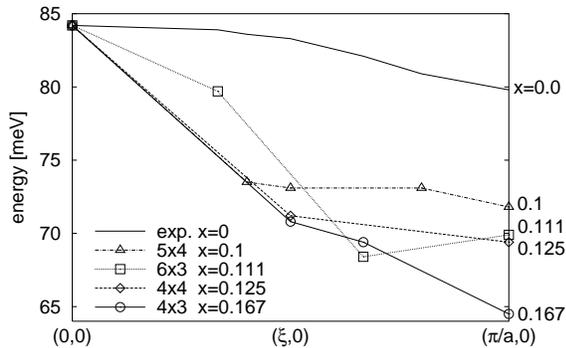}}}}}
\caption[]{\label{fig:2} Phonon dispersion in the (1,0) direction for 
clusters of different sizes. The corresponding doping is indicated.
The figure shows that the softening increases with doping. 
}
\end{figure}

The prefactor of the dominating diagonal term [Eqs. (\ref{eq:3}) and 
(\ref{eq:5})] in the coupling for the half-breathing mode is $\sim$ 
sin$(q_xa/2)$, suggesting a softening $\sim {\rm sin}^2 (q_xa/2)$. 
This behavior is essentially found in the calculations, although the 
softening is stronger for $q_x=\pi/(2a)$ than would be expected from 
this argument.  The prefactor, however, is a factor $\sqrt{2}$ larger 
for the breathing mode [${\bf q}=\pi/a(1,1)$] than for the half-breathing 
mode [${\bf q}= \pi/a(1,0)$], suggesting twice as large a softening for 
the breathing mode. Actually, the softening is larger for the 
half-breathing mode, since it couples to excitations at lower energies.

Figure \ref{fig:2} shows results in the (1,0) direction for different 
cluster sizes. Since all clusters have two holes, the change of 
cluster size changes the doping. The softening increases with 
doping $x$, e.g., at the zone boundary, the softening is 10$\%$ for 
$x=0.1$ and 19$\%$ for $x=0.167$. This is also observed experimentally, 
e.g, for La$_{2-x}$Sr$_x$CuO$_4$ the softening is about 10$\%$ for 
$x=0.10$ \cite{Pintschovius1} and about 14$\%$ for 
$x=0.15$ \cite{Pintschovius2}. For the breathing mode [${\bf q}=\pi/a(1,1)$],
we obtain the softening 8$\%$ ($x=0.125$), 
compared with the experimental results 3$\%$ ($x=0.1$) and 8$\%$ 
($x=0.15$) \cite{Pintschovius3}.

The calculations show that there is a strong coupling to the
half-breathing mode, where two O atoms move towards the Cu atom in 
between.  At the zone boundary, the Cu atoms do not move. Towards the
zone center, however, there is a substantial movement of Cu. 
Normalization of the polarization vectors then leads to a reduction
of the O polarization vectors. Completeness requires that the missing 
O weight is transferred to other modes. In a two-spring model the 
weight goes to an acoustic mode. Because of its small frequency, this mode 
is then softened far more [almost 50$\%$ for ${\bf q}=\pi/(2a)(1,0)$]  
than observed experimentally. 

\begin{figure}[]
\centerline{
{\rotatebox{0}{\resizebox{7cm}{!}{\includegraphics {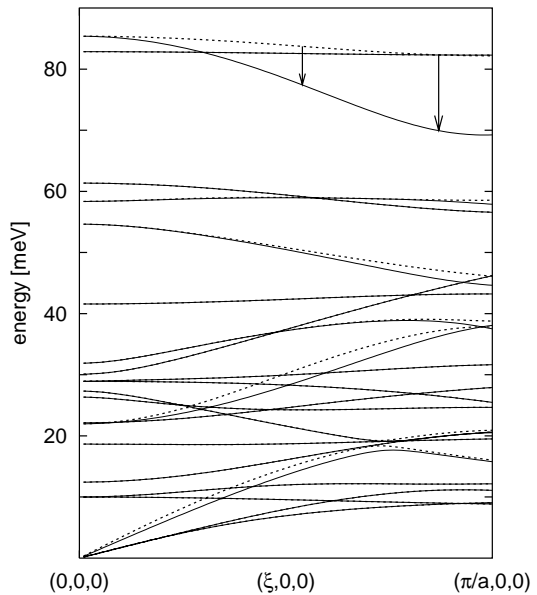}}}}}
\caption[]{\label{fig:3} Phonon dispersion in the (1,0,0) direction for 
a shell model. The dashed curves show results for the undoped
system \cite{Pintschovius1} and the full curves results with the
extra O-O spring describing the O atom coupling to the Zhang-Rice
singlets. The arrows indicate the strong softening of the half-breathing
mode, while other modes are not changed very much. 
}
\end{figure}

To address this, we have used a more realistic shell 
model \cite{Pintschovius1} for obtaining eigenvectors. This model 
gives almost exactly the same eigenvectors for the half-breathing 
mode as our two-spring model. The ``missing'' O weight towards the 
zone center, however, is distributed over several modes, and 
the softening of a given mode is weaker. For instance, the longitudinal
acoustic ${\bf q}=\pi/(2a)(1,0)$ phonon is softened by about 25$\%$. 
Although smaller than in the two-spring model, the softening is still
too large. It is, however, further reduced by the repulsion from 
lower-lying modes of the same symmetry. 

To study this,  we have modified
the shell model \cite{Pintschovius1} to take the electron-phonon
interaction into account. The movement of two O atoms towards a Cu 
atom leads to a lowering of the Zhang-Rice singlet energy. The system 
can take advantage of this by transferring a singlet to such a site. 
This is approximately described by introducing a spring with a {\it 
negative} spring constant, $\kappa=-3$ eV/\AA$^2$, between two O atoms 
on opposite sides of a Cu atom. A similar term was used to describe 
La$_{2-x}$Sr$_x$NiO$_4$ \cite{Pintschovius4,Tranquada}. The present work
gives justification for such a spring. Figure \ref{fig:3} compares results
of the shell model with and without the additional spring. Apart from 
the half-breathing mode, no  mode is strongly softened by the 
new spring \cite{spring}. The $t$-$J$ model thus correctly softens 
the half-breathing mode, without introducing unphysical softening 
of other modes.   

An LDA calculation for the frequency of the half-breathing mode in 
(doped) YBa$_2$Cu$_3$O$_7$ found good agreement with experiment \cite{Bohnen},
although the very small calculated electron-phonon coupling would
suggest a weak doping-dependence. Since LDA cannot describe
the insulating undoped system, the implications for the doping induced
softening are unclear.

The half-breathing and (in particular) breathing modes have 
unfavorable ${\bf q}$ dependences for $d$-wave pairing. 
The coupling to these modes is not expected to explain
superconductivity in Eliashberg-like theories.

Recent photoemission experiments suggest  strong coupling to a phonon
at 40 meV, perhaps a $B_{1g}$ buckling optical phonon,
i.e., a phonon where the in-plane O atoms have $c$-axis out-of-phase displacements \cite{Shen}. 
For symmetry reasons, this mode couples only to second order for a single 
flat CuO$_2$ plane, due to the hopping between the Cu $d_{x^2-y^2}$ and 
O $p_z$ orbitals. Such terms, however, have a  small prefactor, and 
the coupling constant for the on-site term is more than an 
order of magnitude smaller than for the half-breathing mode \cite{qdep}. 
The planes may have local static bucklings, allowing coupling to linear 
order. Assuming a 0.2 \AA \ (6$^{\circ}$) buckling, we 
find that the coupling constant is still an order of magnitude smaller 
for the buckling mode. It seems that one would have to go beyond a single 
layer $t$-$J$ model to obtain a strong coupling to the buckling mode.

We have found that the distance dependence of $t_{pd}$ is substantially
more important than that of $U_{pd}$ for the electron-phonon 
interaction. Nevertheless, the interference effects cannot  be neglected.
The $t$-$J$ model with phonons describes the strong renormalization
of the half-breathing mode, a weaker renormalization of the
breathing mode and no anomalies in other modes.

We thank O. Jepsen for providing unpublished band structure
calculations and O.K. Andersen, P. Horsch, N. Nagaosa and 
Z.-X. Shen for many useful discussions.


\begin{thebibliography}{*}
\bibitem{qdep}The buckling (half-breathing) mode has the strongest
 coupling at the zone  center (boundary).
\end{thebibliography}
\end{document}